\documentclass[twocolumn,trackchanges]{aastex701}

\newcounter{magicrownumbersCP}
\newcounter{magicrownumbersvetoed}

\usepackage{amsthm}

\usepackage{needspace}
\usepackage{bbding}
\usepackage{bm}
\usepackage{microtype}
\usepackage{graphicx,subfigure}
\usepackage{diagbox}
\usepackage{enumitem}
\usepackage{amsmath}
\usepackage{multirow}
\usepackage{makecell}
\usepackage[linesnumbered, ruled, vlined]{algorithm2e}
\usepackage{array}
\usepackage{xcolor}
\usepackage[flushleft, para]{threeparttable}
\usepackage{tabularx}
\usepackage{booktabs}
\usepackage{scrextend}
\usepackage{comment}
\usepackage{hyperref}
\usepackage{etoolbox}

\newcolumntype{Y}{>{\centering\arraybackslash}X}

\newcolumntype{C}[1]{>{\centering\arraybackslash}p{#1}}

\setitemize{leftmargin=10pt,noitemsep,topsep=0pt,parsep=0pt,partopsep=0pt}

\makeatletter
\def\@fnsymbol#1{\ensuremath{
\ifcase#1\or *\or \dagger\or \ddagger\or\
\mathsection\or \mathparagraph\or \|\or **\or \dagger\dagger
\or \ddagger\ddagger \else\@ctrerr\fi
}}

\makeatother

\newcommand{\ssymbol}[1]{$^{\color{red}\@fnsymbol{#1}}$}

\SetKwInput{KwInput}{Input}
\SetKwInput{KwOutput}{Output}

\newcolumntype{P}[1]{>{\hspace{0pt}}p{#1}}

\newcommand{\kepler}{Kepler}
\newcommand{\KeplerMission}{Kepler Mission}
\newcommand{\TESSMission}{TESS Mission}
 
\newcommand{\ExoMiner}{\texttt{ExoMiner}} 
\newcommand{\ExoMinerplusplus}{\texttt{ExoMiner++}}  
\newcommand{\ExoMinerplusplusFFIVetter}{\texttt{ExoMiner++ 2.0}}  
\newcommand{\AstroNet}{\texttt{AstroNet}}  
\newcommand{\AstroNetVetting}{\texttt{AstroNet-Vetting}}

\begin{document}

\title{ExoMiner++ 2.0:  Vetting TESS Full-Frame Image Transit Signals}

\correspondingauthor{Hamed Valizadegan}
\email{hamed.valizadegan@nasa.gov}

\author[0000-0002-2188-0807]{Miguel J. S. Martinho}
\affiliation{KBR, Inc., Mountain View, CA 94043, USA}
\affiliation{NASA Ames Research Center (NASA ARC), Moffett Field, CA 94035, USA}
\email{miguel.martinho@nasa.gov}

\author[0000-0001-6732-0840]{Hamed Valizadegan}
\affiliation{KBR, Inc., Mountain View, CA 94043, USA}
\affiliation{NASA Ames Research Center (NASA ARC), Moffett Field, CA 94035, USA}
\email{hamed.valizadegan@nasa.gov}

\author[0000-0002-4715-9460]{Jon M. Jenkins}
\affiliation{NASA Ames Research Center (NASA ARC), Moffett Field, CA 94035, USA}
\email{jon.jenkins@nasa.gov}

\author[0000-0003-1963-9616]{Douglas A. Caldwell}
\affiliation{The SETI Institute, Mountain View, CA  94043, USA}
\affiliation{NASA Ames Research Center (NASA ARC), Moffett Field, CA 94035, USA}
\email{douglas.caldwell@nasa.gov}

\author[0000-0002-6778-7552]{Joseph D. Twicken}
\affiliation{The SETI Institute, Mountain View, CA  94043, USA}
\affiliation{NASA Ames Research Center (NASA ARC), Moffett Field, CA 94035, USA}
\email{jtwicken@seti.org}

\author[0000-0003-2053-0749]{Ben Tofflemire}
\affiliation{The SETI Institute, Mountain View, CA  94043, USA}
\affiliation{NASA Ames Research Center (NASA ARC), Moffett Field, CA 94035, USA}
\email{ben.tofflemire@nasa.gov}

\author[0000-0001-8019-6661]{Marziye Jafariyazani}
\affiliation{The SETI Institute, Mountain View, CA  94043, USA}
\affiliation{NASA Ames Research Center (NASA ARC), Moffett Field, CA 94035, USA}
\email{marziye.jafariyazani@nasa.gov}

\submitjournal{AJ}


\begin{abstract}
The \textit{Transiting Exoplanet Survey Satellite} (TESS) Full-Frame Images (FFIs) provide photometric time series for millions of stars, enabling transit searches beyond the limited set of pre-selected 2-minute targets. However, FFIs present additional challenges for transit identification and vetting. In this work, we apply \ExoMinerplusplusFFIVetter, an adaptation of the \ExoMinerplusplus\ framework originally developed for TESS 2-minute data, to SPOC FFI light curves. The model is used to perform large-scale planet versus non-planet classification of Threshold Crossing Events across the sectors analyzed in this study. We construct a uniform vetting catalog of all evaluated signals and assess model performance under different observing conditions. We also introduce a novel image-based representation encoding of the relative location and brightness of known neighboring TIC stars, improving the model's ability to identify off-target transit sources and contamination-driven false positives. We find that \ExoMinerplusplusFFIVetter\ generalizes effectively to the FFI domain with PR AUC$\approx0.95$ under multi-source training, demonstrating effective generalization beyond the canonical 2-minute domain, and providing robust discrimination between planetary signals, astrophysical false positives, and instrumental artifacts. This work extends the applicability of \ExoMinerplusplus\ to the full TESS dataset and supports future population studies and follow-up prioritization.
\end{abstract}

\section{Introduction}\label{introduction}

The Transiting Exoplanet Survey Satellite~\citep[TESS;][]{ricker2015TESS} mission has conducted a high-precision and wide photometric survey of the sky over the past seven years and unearthed more than 7,500 planet candidates while enabling the confirmation of more than six hundred new planets. Now into its third Extended Mission (EM3), TESS continues to expand its sky survey through a series of 27-day observation cycles, each covering a $24^\circ\times96^\circ$ field of view. The observatory collects data using mainly two different modes: 2-minute sampling of up to 20,000 targets per sector and full-frame images (FFI)\footnote{Target pixel files and light curves are generated for up to 3000 20-second targets  per sector since sector 27, but these are not searched for transiting planet signatures. All 20-second targets are also observed at the 2-minute cadence.}. For the Prime Mission, the FFI cadence was set to 30 minutes, decreasing to 10 minutes for the First Extended Mission, and to 200 seconds for the following EM2 and EM3. By reducing the cadence for FFI data, the extended missions provide increased sensitivity to short duration transits for more targets and enabled the study of transient stellar events. The TESS Science Processing Operations Center~\citep[SPOC;][]{Jenkins2016SPIE} pipeline has been used since the start of the \TESSMission\ to process 2-minute data to generate data products that include target pixel and light curve files for the different sectors, and the results of transiting planet searches that yield catalogs of Threshold Crossing Events (TCEs), periodic dimming events that might be consistent with planetary transits. The SPOC also produces reports and time series for TCEs that can be used towards vetting and validation efforts. For FFI data, the SPOC pipeline has been used to calibrate FFIs and assign world-coordinate system information to the FFI data delivered to the Mikulski Archive for Space Telescopes\footnote{\url{https://archive.stsci.edu/}} (MAST). Starting with Year 2, the SPOC began processing 30-minute targets selected from the FFIs to create calibrated target pixel files and light curves for up to 160,000 targets per sector in the TESS Northern hemisphere, and has continued to do so, with results released up to Year 6 (as of 11 December 2025, through Sector 78). Furthermore, since Sector 36 (Year 3), the SPOC has been running transiting planet searches on the hundreds of thousands of targets selected, yielding Data Validation~\citep[DV;][]{Twicken_2018_DV, Li2019KeplerDataValidation2} products similar to those created for the 2-minute cadence data (for those targets with at least one detected TCE). All of these data products have been delivered to MAST as High-Level Science Products\footnote{\url{https://archive.stsci.edu/hlsp/tess-spoc/}} (HLSP) for FFI targets from Sectors 36 onwards~\citep{caldwell2020tess}.

The FFI transit searches on a larger population of target stars increase significantly the number of transiting signals that have been detected by data processing pipelines such as SPOC for the \TESSMission. The vast volumes of TCEs produced by these pipelines require vetting systems that are automated, scalable, and capable of vetting faint transit signals with a minimal false positive rate. As demonstrated by the previous works~\citep{shallue_2018, armstrong-2020-exoplanet, Valizadegan_2022_ExoMiner, Valizadegan_2023_Multiplicity, Valizadegan_2025}, machine learning methods are particularly well suited for these requirements, as they can identify patterns in a data-driven manner to efficiently vet hundreds of thousands of transiting signals. This allows the exoplanet community to focus their time and resources on a smaller set of candidates by excluding false positives from the initial population generated by transit search runs. 


In this work, we build upon our previous efforts in vetting TESS SPOC 2-minute TCEs~\citep{Valizadegan_2025}, and extend these efforts to the FFI planet searches. More concretely, we present our methodology for vetting TCEs generated from SPOC planet search runs conducted on FFI data. We improve the architecture of \ExoMinerplusplus, the deep learning model used in~\citep{Valizadegan_2025}, and train this model for finding planet candidates among the hundreds of thousands of these FFI transiting events, with the goal of providing catalogs that enable focused follow-up observation efforts in a smaller and cleaner set of candidates. These catalogs significantly speed up the exoplanet community's efforts to confirm and validate planetary candidates. Candidate identification is supported by both the TESS Science Office systematic vetting process \citep{guerrero2021TOI}, which produces the official TOI list, and community-driven projects, e.g.~\cite{eisner2021planet}, that contribute additional candidates to the CTOI catalog.

The paper is structured as follows: we begin by detailing the data used for transit signal classification and the modifications to the \ExoMinerplusplus\ architecture in Sections~\ref{sec:data_preprocessing} and~\ref{sec:model}, including label sources and sector run usage. Section~\ref{sec:experimental_setup} outlines our experimental setup for model learning and evaluation, where we specifically investigate the model's performance on FFI data compared to 2-minute data, and assess the impact of supplying neighbor location and brightness information as inputs to the model. The paper concludes in Section~\ref{sec:vetting_catalog}, which presents the resulting vetting catalog from applying the trained model to FFI TCEs, alongside an analysis of the catalog's characteristics based on parameters like orbital period and planet radius.

\section{Related Work}


The development of automated vetting frameworks of transit signals followed the fast increase in the data volume and throughput of exoplanet survey missions in the last decade, with \KeplerMission ~\citep{borucki2010kepler} being a tipping point in that domain. The \KeplerMission\ provided catalogs of hundreds of thousands of transit signal detections that have made it the most prolific exoplanet survey mission to date, with its data products still actively explored for applications that include, but also go beyond, exoplanet discovery. The ongoing \TESSMission\ follows in the steps of \kepler\ and continues to deliver thousands of transit signal events per sector, with more in the years to come. Future missions such as NASA's Roman Space Telescope~\citep{green2012wide} and ESA's PLAnetary Transits and Oscillations~\citep[PLATO;][]{rauer2010plato} pick up such efforts to use transit photometry to unearth new extrasolar planets and fill in the gaps in our understanding of planet population statistics. It is expected that these upcoming missions will create large volumes of data to be added to the already large set being delivered by TESS. Thus, it becomes crucial to develop accurate and efficient automated methods that can sift through these large datasets and uncover planet candidates suitable for follow-up observations. Several transit signal vetters have been developed throughout the last decade, and include if-then rule-based methods~\citep{Coughlin_2016_robovetter, tec}, statistical frameworks~\citep{Morton_2011_Vespa, Morton_2012_vespa, Morton-2016-vespa, Giacalone_2020_TRICERATOPS}, and more recently machine learning algorithms such as random forests~\citep{Jenkins-Autovetter-2014IAUS, McCauliff_2015} and deep neural networks~\citep{shallue_2018, Ansdell_2018, armstrong-2020-exoplanet, Valizadegan_2022_ExoMiner, Valizadegan_2023_Multiplicity}.

Regarding the application of transit signal classification frameworks to TESS data, there are a few works that include: (1) \ExoMinerplusplus ~\citep{Valizadegan_2025}, a multi-branch convolutional neural network (CNN) designed to mimic subject matter experts who examine various false positive tests in the reports produced by the SPOC pipeline DV module. \ExoMinerplusplus\ processes several flux time series, periodogram, difference image data, and statistical diagnostics to vet 2-minute cadence TCEs from the SPOC pipeline up to the end of Year 5; (2) \AstroNetVetting\ developed for TESS FFI data~\citep{yu2019identifying}, which is a multi-branch CNN built on the architecture of \AstroNet~\citep{shallue_2018} and adds as input a few more diagnostic tests generated by the Quick Look Pipeline~\citep{2020RNAAS...4..204H, 2020RNAAS...4..206H} from calibrated FFI data for early Sectors 1--6; and (3) the work in~\cite{Tey_2023_astronet}, another modified version of \AstroNet, conduct a triaging of QLP FFI TCEs using data from Years 1 through 3, with the main focus of separating periodic eclipsing signals (which include both planets and non-contact eclipsing binaries) from other astrophysical phenomena and instrumental noise. To the best of our knowledge, this work is the first to incorporate the relative magnitude and location of known neighboring TIC stars, encoded explicitly as an image representation, into a machine learning vetting framework for TESS transit signals.

\section{Data}

As a supervised machine learning model, \ExoMinerplusplus\ required a dataset of TCEs for training, validation, and evaluation. In this section, we describe the catalogs of TCEs that we use to build our dataset, the sources of data and dispositions for the TCEs, and the preprocessing steps conducted to create a dataset of examples. Throughout this work, a TCE is referred to as TESS SPOC \textit{W} TCE TIC \textit{X}-\textit{Y}-S\textit{Z}, where \textit{X}, \textit{Y}, and \textit{Z} represent the TIC ID, SPOC planet number, and SPOC sector run associated with the TCE, respectively. \textit{W} indicates whether the TCE originated from FFI or 2-minute (2-min) data.

\subsection{Data Sources}~\label{sec:data_sources}

In this work, we use the light curves and other data products generated by the transit searches performed by the SPOC pipeline~\citep{twickenSDPDD2020} for both TESS 2-minute and FFI data, which are available at MAST (the latter as HLSP data). Specifically, we use the light curve FITS files to extract the Presearch Data Conditioning Simple Aperture Photometry~\citep[PDCSAP;][]{2012PASP..124..985S, 2014PASP..126..100S, 2012PASP..124.1000S} flux and the flux-weighted centroid motion time series for each target in any given sector that is associated with a TCE in our dataset, and use the DV XML files to extract the ephemerides, statistics, fit parameters, and difference image data for each TCE. 

For our target population, we use TIC-8~\citep{2019AJ....158..138S} as the source catalog for stellar parameters and Gaia DR2~\citep{2018A&A...616A...1G} as the source for Renormalized Unit Weight Error~\citep[RUWE;][]{LL:LL-124} values.

\subsection{Sector Runs}~\label{sec:sector_runs}

The dataset of TCEs used in this work can be split into two main sets: the SPOC TCEs generated by runs using 2-minute data (from now on referred to as 2-minute TCEs), and those generated by the SPOC runs on the FFI data (from now on referred to as FFI TCEs). As with any SPOC DV product, those TCE datasets are available at MAST. The following sector runs were considered for each one of these two populations:

\begin{itemize}
    \item 2-minute TCEs: single-sector runs from Sector 1 through Sector 88, and twenty multisector runs that include Sectors 1--2, 1--6, 1--9, 1--13, 1--36, 1--39, 1--46, 1--65, 1--69, 2--72, 14--19, 14--23, 14--26, 14--41, 14--50, 14--55, 14--60, 14--78, 42--43, and 42--46.
    \item FFI TCEs: single-sector runs from Sector 36 through Sector 72, and one multisector run for Sectors 56--69.
\end{itemize}

\subsection{Label Sources and Assignment}
\label{sec:label-assignment}
The training and evaluation of our model requires a dataset of labeled TCEs, which entails first finding dispositions based on catalogs of confirmed planets and different types of false positives (e.g.,\ eclipsing binaries, background objects, stellar variability, and instrumental noise); second, matching those events to TCEs through ephemeris matching, following the procedure outlined in~\citep{Twicken_2018_DV}; and third, mapping those dispositions to a set of labels that are consistent and accurate. It is crucial that we minimize the amount of label noise in our dataset so we can have a high-quality labeled dataset for training and benchmarking our model. 

We follow the procedure described in~\citep{Valizadegan_2025} and use the Exoplanet Follow-up Observing Program (ExoFOP) TOI catalog\footnote{\url{https://exofop.ipac.caltech.edu/tess/} downloaded September 22, 2025}, Prsa's TESS eclipsing binary (EB) catalog~\citep{Prsa_2022-EB-catalog}, and TESS-ExoClass~\citep[TEC;][]{tec} tables for SPOC 2-minute and FFI TCEs (for 2-minute TCEs, we use TEC results for all single-sector runs from Sector 1 through Sector 41; for FFI TCEs, we use results from single-sector runs in range 40--72). From the ExoFOP TOI catalog, we use the confirmed planet (CP), known planet (KP), and false positive (FP) dispositioned by the TFOP Working Group (TFOPWG). We consider the KP and CP ExoFOP TOIs that are known brown dwarfs (BDs) based on TFOP Subgroup 1 (SG1) TOI catalog photometric dispositions as false positives and label them as such. TEC flux triage results are used to label TCEs as non-transiting phenomenon (NTP) and Prsa's TESS EB catalog was used to obtain EB TCEs. 
At the end of this labeling process, the TCEs that lack a label are labeled as unknown (UNK). To obtain a binary label, TCEs dispositioned as CP and KP are labeled `planets' and those dispositioned as BD, EB (i.e.,\ Prsa's EBs), FP, and NTP are labeled `non-planets'. Refer to~\cite{Valizadegan_2025} for the details of this labeling procedure. 

\needspace{3\baselineskip} 
\subsection{Data Preprocessing}~\label{sec:data_preprocessing}

The data described in Section~\ref{sec:data_sources} are preprocessed following the steps in~\citep{Valizadegan_2025} to generate the examples for the TCEs in our datasets. The preprocessing pipeline involves the ephemeris-based (i.e.,\ period, transit duration and epoch of first transit) phase-folding and binning of flux (e.g.,\ primary, secondary, and odd/even events) and flux-weighted centroid motion time series, periodogram computation, and the extraction and preprocessing of in-transit, out-of-transit, and difference flux image data for each TCE. Other scalar features such as the target stellar parameters and TCE fit parameters and statistics computed in the DV module of the SPOC pipeline are used as ancillary features to those higher-dimensionality data. The reader is invited to consult~\citep{Valizadegan_2025, Valizadegan_2022_ExoMiner, Twicken_2018_DV, twickenSDPDD2020} for a comprehensive discussion on the data preprocessing and on the SPOC pipeline data products. We made the following changes to the preprocessing pipeline described in~\citep{Valizadegan_2025}:

\begin{itemize}
    \item \textbf{Discontinued preprocessing of momentum dump events}: following the conclusions in~\citep{Valizadegan_2025}, we chose not to use the momentum dump quality flags as input to our model, as this information was not found to be useful for distinguishing planet TCEs from their non-planet counterparts.
    
    \item \textbf{Encoding of known neighboring stars' location and magnitude}: the model described in~\citep{Valizadegan_2025} was provided with information regarding the target location along with the out-of-transit and difference flux images. This set of information enables the model to understand whether there is an offset of the transit source relative to the location of the target (by mapping the target's TIC coordinates to the CCD pixel frame) and to the out-of-transit centroid. A larger offset might be indicative that the transiting event occurs on a background object. However, the model was not aware of the existence of neighboring stars and their relative magnitudes that show up in the target mask pixels. Without this information, the model is completely blind to scenarios like crowded fields where there can be other stars relatively bright that can be responsible for the transit observed. Thus, we extract information about the location and the magnitude of neighboring stars for each target, and preprocess these information into an image (as an extra input channel in the `Difference Image' branch) and feed it to the model. Section~\ref{sec:neighbors_img_preprocessing} describes the preprocessing of the location and magnitude of neighboring stars to create an image that encodes such information.
    
    \item \textbf{Increased image resolution for difference image data}:~\cite{Valizadegan_2025} sampled the difference image data using a factor of 3 so the target sub-pixel location could be provided at a higher resolution than pixel-level (e.g.,\ images with an original size of $11\times11 \, \text{px}$ were resampled to a final size of $33\times\ 33 \, \text{px}$ so each pixel in the original image was mapped to a $3\times3 \, \text{px}$ region). Since photocenter offsets of 4–6 arcseconds are generally considered reliable \citep{2025ApJS..279...50K}, we adopt a factor of 5 here, which results in image dimensions of $55 \times 55 \,\text{px}$. This adjustment provides the model with finer spatial resolution for locating the target and nearby stars.
    
    \item \textbf{Excluded target image from the `Difference Image' branch}:~\cite{Valizadegan_2025} used an image to provide the location of the target star to the model. Given that this image only includes the target location information, it is very sparse and difficult for the model to learn from. In this work, instead, we encode the target location by centering the difference image data on the pixel containing the target star. This removes the need to have an extra image channel encoding the target location. 

    \item \textbf{Added Signal-to-Noise Ratio (SNR) flux image to `Difference Image' branch}: we computed, preprocessed, and added the SNR flux image as another input channel for the difference image data. The SNR flux image provides information on the significance of the estimated flux difference between out-of-transit and in-transit cadences. It is computed as the difference image flux divided by the uncertainty in each pixel~\citep{Twicken_2018_DV}. 
\end{itemize}

\begin{figure*}[htb!]
\begin{center}
\centerline{\includegraphics[width=\textwidth]{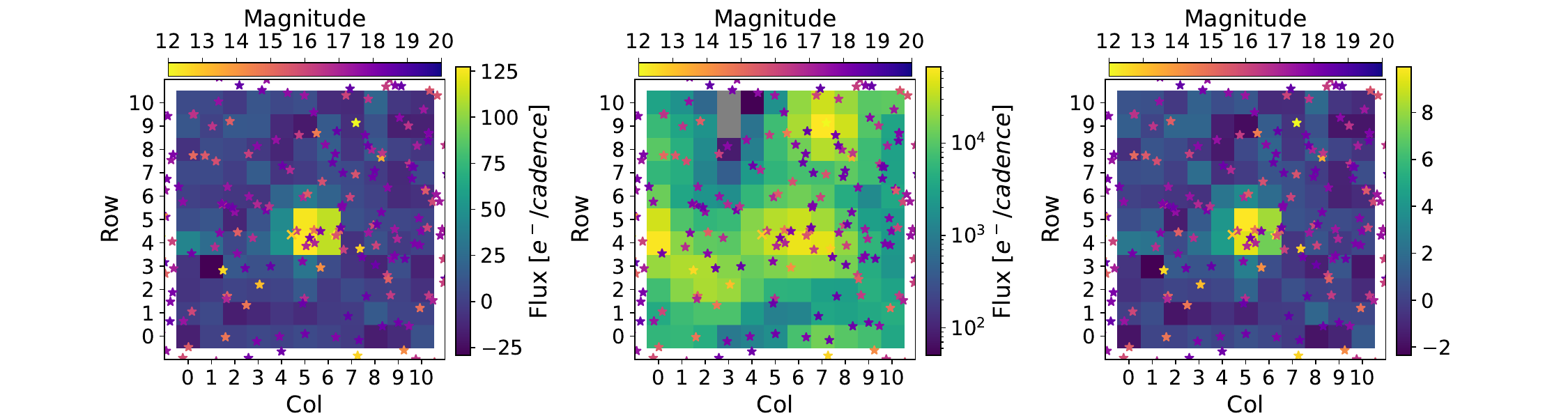}}
\vspace{0.05in}
\caption{From left to right, the panels display the difference, out-of-transit, and SNR fluxes for TESS SPOC 2-minute TCE TIC 83053699-1-S57 (TOI 4002.01). The neighboring stars and target's TIC-8 coordinates mapped to the CCD frame are identified as stars and as a cross, respectively. Their color encodes the stars' $T_{mag}$.}
\label{fig:search_neighbors_example}
\end{center}
\end{figure*}

\begin{figure}[htb!]
\begin{center}
\includegraphics[width=\linewidth]{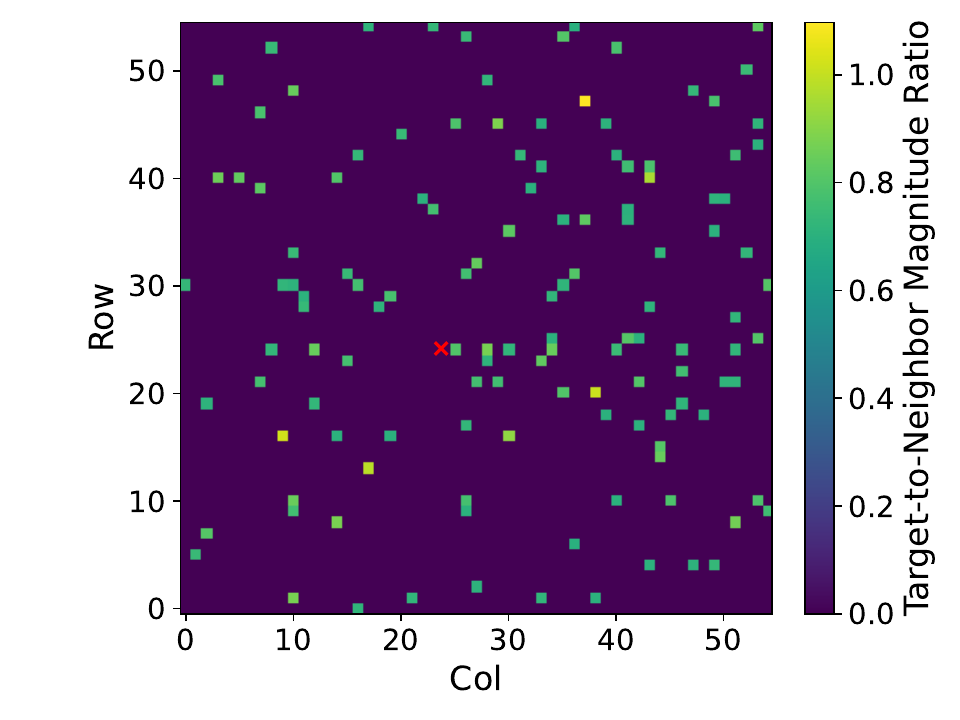}
\caption{Preprocessed and unnormalized neighboring stars image for TESS SPOC 2-minute TCE TIC 83053699-1-S57 (TOI 4002.01). The target's TIC-8 position is shown as a red cross.}
\label{fig:preprocessed_neighbors_img_example}
\end{center}
\end{figure}


\subsubsection{Encoding Neighboring Stars' Location and Magnitude}
\label{sec:neighbors_img_preprocessing}

To encode the location and magnitude of neighboring stars, we queried the TIC-8 catalog to search for stars in a search radius of $168 \, \text{arcsec}$ ($\approx 8 \, \text{px}$) around the target's TIC coordinates. For each target and sector run of interest, we used the World Coordinate System (WCS) stored in the corresponding light curve file to map the celestial coordinates of the neighboring stars into pixel coordinates. Figure~\ref{fig:search_neighbors_example} shows an example of such search for TIC 83053699 in Sector 57. The two images, difference and out-of-transit fluxes for TCE 1 (TOI 4002.01, a false positive attributed to a nearby eclipsing binary), show the location of the neighboring stars that were found in a search radius of $168 \, \text{arcsec}$, colored as a function of their magnitude value obtained from TIC-8. 

With this catalog of neighboring stars for each target-sector pair (i.e.,\ a target star observed in a given sector), we create the neighbors images for each TCE. We start by excluding multiple different types of neighboring stars: 
\begin{itemize}
    \item Neighbors whose brightness indicates that even a full eclipse could not account for the observed transit depth are excluded. Let $T_{mag}^n$ and $T_{mag}^t$ be the TESS magnitude of a neighboring star and the TESS magnitude of the target, respectively, and let $\delta$ be the estimated fractional transit depth of the TCE detected for the target. Under the simplifying assumption that the target star dominates the aperture flux ($F_t \gg F_n$), the maximum depth a contaminating neighbor can produce is approximated as $\delta \approx F_n/F_t$. A neighboring star is excluded if the magnitude difference satisfies: $T_{mag}^n - T_{mag}^t >-2.5*log_{10}(\delta)$. This threshold is conservative in the sense that it avoids prematurely excluding potential contaminants. It removes only those that are unequivocally too faint to produce the observed depth, while retaining any neighbor that could potentially be responsible. 
    
    \item All neighboring sources falling within the same pixel (after increasing image resolution by a factor of 5) are excluded except for the brightest one, when multiple neighbors are present, as the brightest source is the most likely contributor to any off-target transit signal.
    
    \item Neighboring stars that fall outside the $11\times11 \, \text{px}$ neighborhood centered on the target pixel are not included. While it is possible that the stars outside of this region are the transit source, encoding their location accurately within the image representation is nontrivial. One potential alternative to include information about stars outside of the patch that still bleed light towards the target neighborhood, would be to extend the edges of the image by $1\,\text{px}$ to encode neighbors outside of the $11\times11 \, \text{px}$ neighborhood, however that also comes with drawbacks in terms of distorting the distance of such neighbors to the target star, and requiring the difference image data to be extended in a similar way. To avoid these challenges, and because it is unlikely that neighbors in this situation are the transit source, we exclude them when creating the neighbors image.
\end{itemize}

\begin{table*}[!t]
\footnotesize
\centering
\caption{TESS 2-minute SPOC TCE and Event Counts.}
\label{table:2min_counts_dataset}
\begin{threeparttable}
\begin{tabularx}{\linewidth}{@{}l|c|cc|cccc@{}}
\cline{2-8}
& Classes & \multicolumn{2}{c|}{Exoplanets} & \multicolumn{4}{c}{Non-planets} \\
\cline{2-8}
& Subclasses & KP & CP & BD & EB & FP & NTP \\
\hline
\multirow{2}{*}{TCE Counts} & Count (Percentage)
& 2{,}587 (4.40\%) & 2{,}852 (4.85\%) & 68 (0.12\%) & 11{,}794 (20.07\%) & 2{,}713 (4.62\%) & 38{,}747 (65.94\%) \\
\cline{2-8}
& Total (Percentage)
& \multicolumn{2}{c|}{5{,}439 (9.26\%)} & \multicolumn{4}{c}{53{,}322 (90.74\%)} \\
\hline\hline
\multirow{2}{*}{Event Counts} & Count (Percentage)
& 546 (13.51\%) & 546 (13.51\%) & 19 (0.47\%) & 2{,}292 (56.72\%) & 638 (15.79\%) & N/A (--\%) \\
\cline{2-8}
& Total (Percentage)
& \multicolumn{2}{c|}{1{,}092 (27.02\%)} & \multicolumn{4}{c}{2{,}949 (72.98\%)} \\
\hline
\end{tabularx}
\end{threeparttable}
\end{table*}

\begin{table*}[!t]
\footnotesize
\centering
\caption{TESS FFI SPOC TCE and Event Counts.}
\label{table:ffi_counts_dataset}
\begin{threeparttable}
\begin{tabularx}{\linewidth}{@{}l|c|cc|cccc@{}}
\cline{2-8}
& Classes & \multicolumn{2}{c|}{Exoplanets} & \multicolumn{4}{c}{Non-planets} \\
\cline{2-8}
& Subclasses & KP & \multicolumn{1}{c|}{CP} & BD & EB & FP & NTP \\
\hline
\multirow{2}{*}{TCE Counts}
& Count (Percentage)
& 806 (1.14\%) & \multicolumn{1}{c|}{1,000 (1.42\%)}
& 30 (0.04\%) & 4,607 (6.54\%) & 1,208 (1.71\%) & 62,787 (89.14\%) \\
\cline{2-8}
& Total (Percentage)
& \multicolumn{2}{c|}{1,806 (2.56\%)}
& \multicolumn{4}{c}{68,632 (97.44\%)} \\
\hline\hline
\multirow{2}{*}{Event Counts}
& Count (Percentage)
& 364 (14.46\%) & \multicolumn{1}{c|}{370 (14.70\%)}
& 14 (0.56\%) & 1,332 (52.92\%) & 437 (17.36\%) & N/A (--\%) \\
\cline{2-8}
& Total (Percentage)
& \multicolumn{2}{c|}{734 (29.16\%)}
& \multicolumn{4}{c}{1,783 (70.84\%)} \\
\hline
\end{tabularx}
\end{threeparttable}
\end{table*}

After excluding those neighboring stars, we set the value of a pixel to the ratio of the target's magnitude to the neighbors' magnitude if there is a neighboring star; otherwise, it is set to zero. With this encoding, pixels with values greater than one represent neighbors that are brighter than the target star, while those with value less than one are dimmer. Figure~\ref{fig:preprocessed_neighbors_img_example} shows the neighbors' image for the same TCE in Figure~\ref{fig:search_neighbors_example} after preprocessing the neighbors data by following these steps. Contrary to the normalization performed for the difference, out-of-transit, and SNR flux images, where we standardize the images using statistics computed from the training set, for the neighbors images we conduct fixed min-max normalization. This choice was motivated by the fact that many pixels have no neighboring stars, resulting in ratio values of zero and consequently an extremely small standard deviation of pixel values, which is required for standardization. Such small standard deviation can cause the non-zero values to become disproportionately large, potentially disrupting model training, hindering convergence, and impairing overall learning. The target-to-neighbor magnitude ratio of all targets in our dataset is not larger than 5, and so we use a factor of 5 to normalize these ratios, ensuring that they are in $[0, 1]$.

\begin{figure}
  \centering
    \subfigure[Target Count]{\label{fig:hist_targets_2min_ffi_tmag_count}\includegraphics[width=0.80\columnwidth]{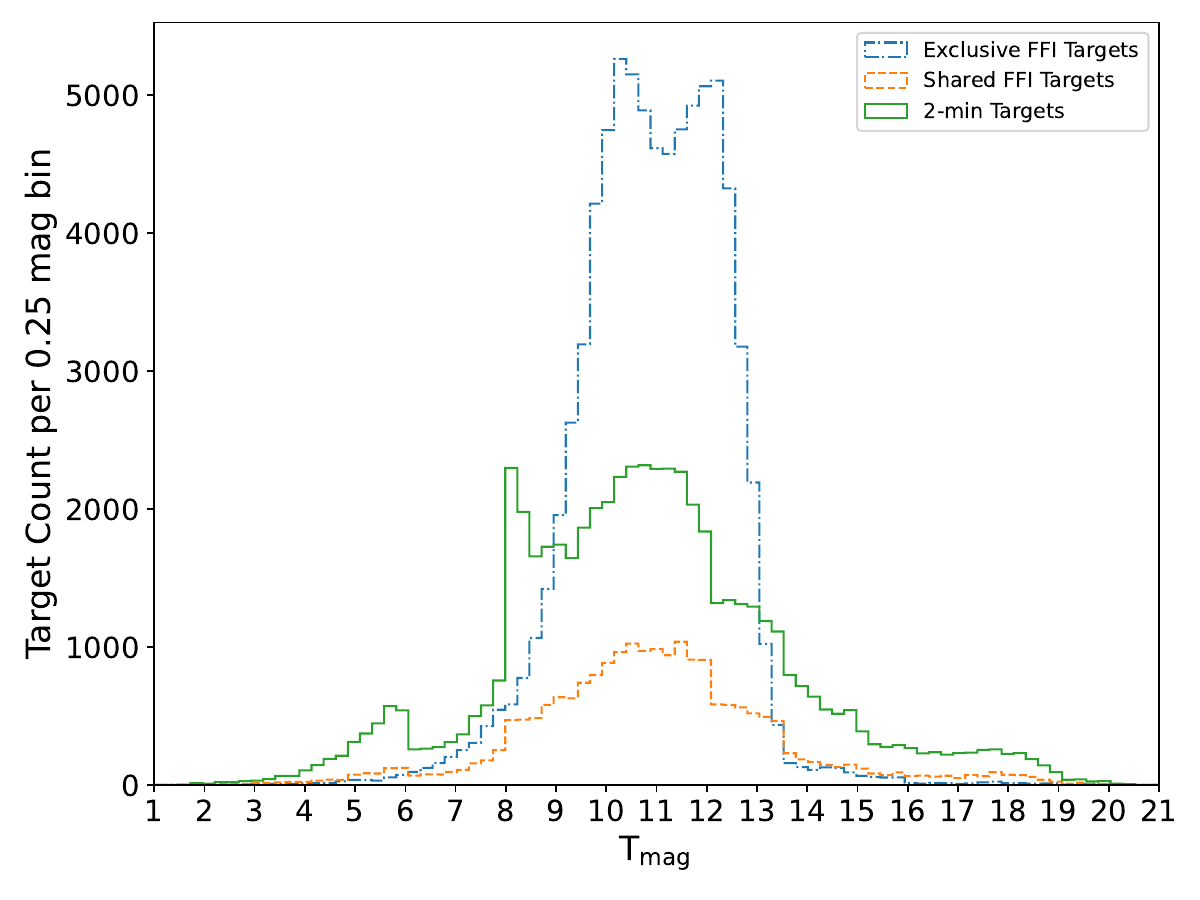}}
    \subfigure[Target Density]{\label{fig:hist_targets_2min_ffi_tmag_density}\includegraphics[width=0.80\columnwidth]{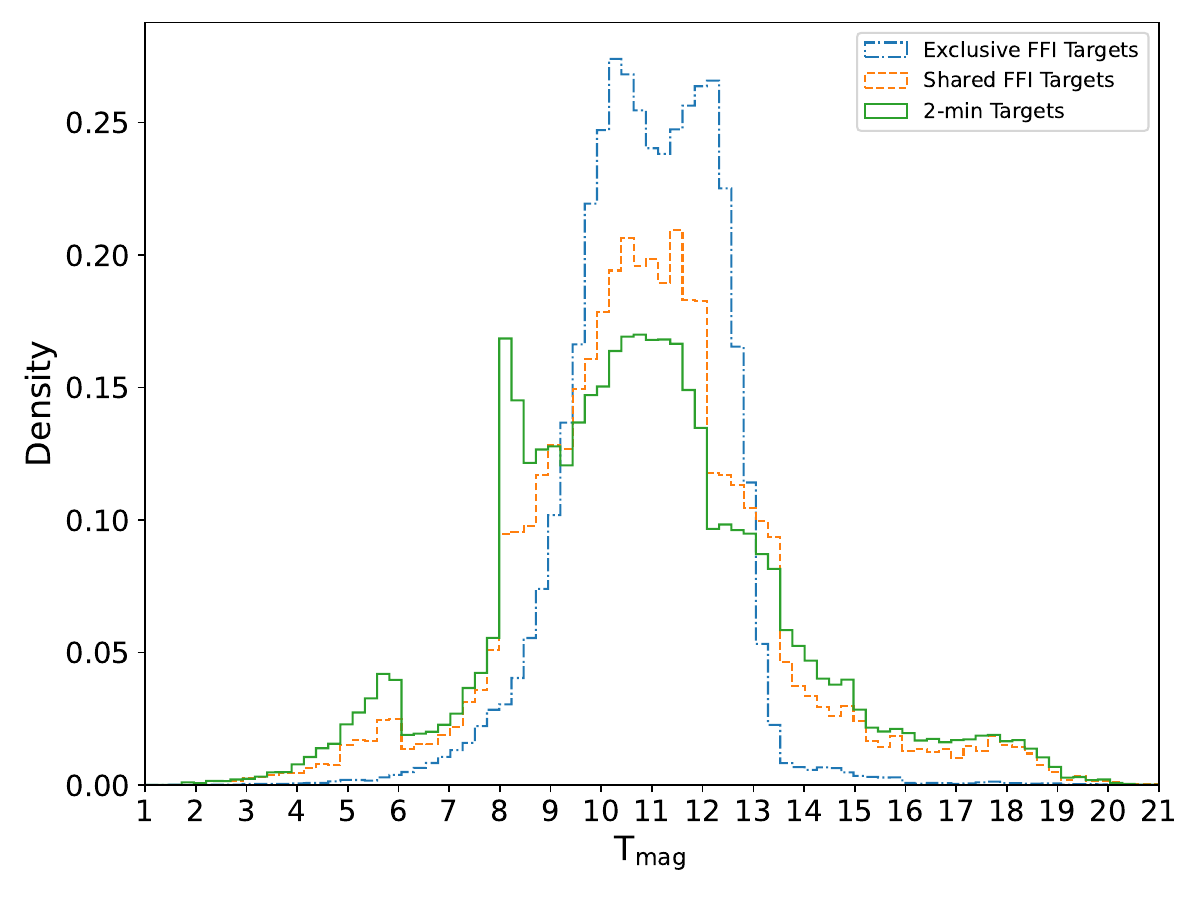}}
\caption{Target population for the 2-minute and FFI TCE datasets as a function of $T_{mag}$.}
\label{fig:hist_targets_2min_ffi_tmag}
\end{figure}

\begin{figure*}[htb!]
\begin{center}
\centerline{\includegraphics[width=\textwidth]{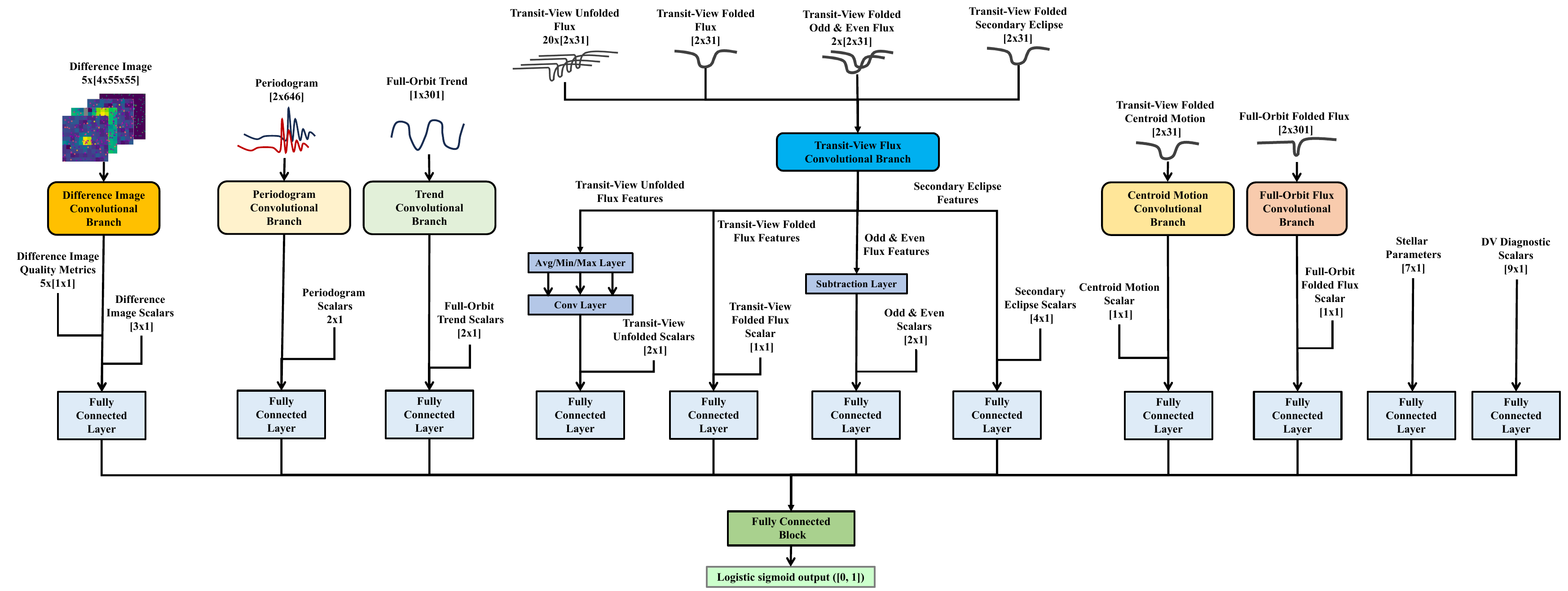}}
\caption{High-level depiction of the new \ExoMiner\ architecture used in this work. The details of this model are described in Section~\ref{sec:model}.}
\label{fig:architecture}
\end{center}
\vskip -0.3in
\end{figure*}

\subsection{Dataset}

Preprocessing the 2-minute and FFI data (i.e.,\ light curves and DV SPOC XML files) resulted in datasets of 268,787 and 222,054 TCEs, respectively. For the FFI data, 70,438 TCEs $(\approx 31.72\%)$ are labeled using the procedure discussed in Section~\ref{sec:label-assignment}, leaving 151,616 as UNK TCEs. 

Tables~\ref{table:2min_counts_dataset} and~\ref{table:ffi_counts_dataset} show the distribution of TCEs and events in the 2-minute and FFI labeled datasets as a function of their binary label (i.e.,\ planet or non-planet class) and disposition (i.e.,\ subclass). Given that more than one TCE can be associated with the same event (e.g.,\ TOI) due to the observation behavior of TESS, these tables also break down the dataset in terms of these unique transiting objects. NTPs are not federated to any transit event given that they are not associated with real astrophysical transiting phenomena. For that reason, we have excluded these TCEs from the event count.

As indicated by the numbers reported in these two tables, the 2-minute dataset contains a larger number of TCEs confirmed as ExoFOP TOIs than the FFI dataset. This difference is primarily due to the more extensive sector coverage, particularly the multisector observations, available for the 2-minute data, which increases the likelihood of detecting longer-period TOIs and planets with smaller radii, and to the smaller number of sectors with HLSP SPOC FFI data products available at this moment.

Figure~\ref{fig:hist_targets_2min_ffi_tmag} shows the distribution of targets in the 2-minute and FFI TCE datasets (including unlabeled TCEs) as a function of $T_{mag}$. The two distributions have a significant overlap, although the FFI target population exhibits a higher concentration of targets around $T_{mag}=[9,13]$, and the 2-minute target population distribution shows a slightly heavier tail for both brighter and dimmer targets. These distinct distributions reflect the differing selection criteria for 2-minute and SPOC FFI targets, highlighting the potential for discovering new planetary candidates within the remaining unvetted TCEs.


\section{Model} \label{sec:model}

The model used in this work, hereafter referred to as {\ExoMinerplusplusFFIVetter}, builds on \ExoMinerplusplus~\citep{Valizadegan_2025}. A high-level depiction of this architecture is shown in Figure~\ref{fig:architecture}, and a detailed version that includes the dimension of the feature maps after each layer can be found on Zenodo~\citep{martinho_2025_17707413}. Several architectural modifications involve changes to the branches responsible for processing specific inputs, as described below:

\begin{itemize}
    \item \textbf{Merged `Unfolded Flux' branch with the other transit-view flux branches}: given that the unfolded binned flux time series shares similarities with the phase-folded and binned flux time series (i.e.,\ they are all transit views of the primary or secondary transits of the TCE), we decided to add the 20 sampled phases of the unfolded flux time series (including the variability time series as the second channel) to the existing four views (primary, secondary, and odd-even) to be processed by the same convolutional branch. The extracted features from the unfolded time series input are then processed separately using the same method described in~\cite{Valizadegan_2025}.
    \item \textbf{Moved transit source offset estimate and uncertainty to `Difference Image' branch}: the transit source computed by the DV module of the SPOC pipeline is estimated from the difference image data. To take that into account, we moved this scalar feature (along with its uncertainty) from the `Centroid Motion' branch to the `Difference Image' branch.
\end{itemize}

\vspace{0.5\baselineskip} 
The remaining changes are related to improvements on the extraction and processing of features through the CNN model and include:

\begin{itemize}
    \item \textbf{Adding global average pooling layers}: we added global average pooling layer instead of basic flattening layer in the end of each convolutional branch. 
    This decreases the number of features at the end of the convolutional branch to the number of channels set for the final convolutional layer, thus leading to a smaller model that is less prone to overfitting. This type of layer has been used in many state-of-the-art (SOTA) CNN architectures such as ResNet~\citep{he2016deep} and Inception~\citep{szegedy2015going}, especially those designed for classification tasks. 
    
    \item \textbf{Replacing max pooling with learned downsampling}: Instead of using a maxpooling layer at the end of each convolutional block, the first convolutional layer in each block applies valid padding with a unit stride, which reduces the spatial resolution similarly to a pooling operation but allows the model to learn the downsampling transformation. This modification is inspired by state-of-the-art architectures, such as more recent variants of ResNet~\citep{he2016deep} and YOLOv5~\citep{jocher2020ultralytics}, which replace pooling operations with learned downsampling so the model is able to learn task-specific downsampling operations, rather than relying on a fixed transformation such as maxpooling.

    \item \textbf{Preserving spatial resolution for difference images}: since detecting small transit source offsets is critical, we did not apply downsampling approaches like strided convolution and max pooling to the `Difference Image' branch. Downsampling can discard fine-grained positional information which are needed to detect subtle offsets. 
    
    \item \textbf{Adding batch normalization and skip connections}: to improve the stability and convergence of the models during training, a batch normalization layer~\citep{ioffe2015batch} was introduced after each convolutional layer. Additionally, a residual block~\citep{he2016deep} was added to the output of each convolutional block to provide better gradient flow and allow feature reuse.
    
    \item \textbf{Adding layer normalization}: to balance the contribution from all branches in the model, we applied a layer normalization layer~\citep{ba2016layer} after concatenating the features extracted from all branches. These layers stabilize training by normalizing the activations across features. Fully connected (FC) layers are able to learn and adjust the weights to compensate for scale differences in the features; however, by applying these layers, we reduce the risk of one branch's features overshadowing the others due to their raw magnitude, since features with larger magnitudes produce larger gradients that can dominate the updates during training. To further improve stability during training and reduce overfitting, we added one normalization layer after each FC layer in the large classification head.
\end{itemize}

\begin{table*}[!t]
\footnotesize
\centering
\caption{Binary classification performance of \ExoMiner\ models for the test sets of the 5-fold cross validation experiment using multi-source data from both 2-minute and FFI sources. Precision, recall, and accuracy are computed at a classification threshold of 0.5. Mean and standard deviation estimates are computed across the CV iterations. The best performer is highlighted in bold. Differences are not claimed to be statistically significant unless stated otherwise.}
\label{table:cv_binary_results}

\setlength{\tabcolsep}{4pt} 
\renewcommand{\arraystretch}{1.2} 
\begin{tabular*}{\textwidth}{@{\extracolsep{\fill}} l | >{\centering\arraybackslash}p{2.5cm} | >{\centering\arraybackslash}p{1.8cm} | >{\centering\arraybackslash}p{1.8cm} | >{\centering\arraybackslash}p{1.8cm} | >{\centering\arraybackslash}p{2.8cm} | >{\centering\arraybackslash}p{2.8cm} @{}}
\hline
Model & Precision \& Recall & PR AUC & ROC AUC & Accuracy & Precision @ 95\% Recall & Recall @ 95\% Precision \\
\hline
\ExoMinerplusplus
  & $0.913\pm0.027$ \& $\mathbf{0.912\pm0.026}$ & $0.962\pm0.012$ & $\mathbf0.996\pm0.002$
  & $0.990\pm0.002$ & $0.872\pm0.015$ & $0.798\pm0.113$ \\
\ExoMinerplusplusFFIVetter
  & $0.922\pm0.019$ \& $0.865\pm0.019$ & $0.958\pm0.006$ & $0.995\pm0.002$
  & $0.988\pm0.002$ & $0.832\pm0.026$ &  $0.751\pm0.074$ \\
\hline
\end{tabular*}
\end{table*}

\begin{table*}[!t]
\footnotesize
\centering
\caption{Subclasses recall at classification threshold of 0.5 for the test sets of the 5-fold cross validation experiment using multi-source data from both 2-minute and FFI sources. Mean and standard deviation estimates are computed across the CV iterations. The best performer is highlighted in bold. Differences are not claimed to be statistically significant unless stated otherwise.}
\begin{threeparttable}
\setlength{\tabcolsep}{4pt} 
\renewcommand{\arraystretch}{1.2} 
\begin{tabular*}{\textwidth}{@{\extracolsep{\fill}} >{\centering\arraybackslash}p{3.5cm} | >{\centering\arraybackslash}p{2.0cm} | >{\centering\arraybackslash}p{2.0cm} | >{\centering\arraybackslash}p{2.0cm} | >{\centering\arraybackslash}p{2.0cm} | >{\centering\arraybackslash}p{2.0cm} | >{\centering\arraybackslash}p{2.0cm} @{}}
\toprule
\diagbox[width=3.5cm]{Model}{Recall} & KP & CP & BD\tablenotemark{a} & EB & FP & NTP \\
\midrule
\ExoMinerplusplus & $0.915\pm 0.022$ & $\mathbf{0.906\pm 0.045}$ & $0.686\pm 0.188$ & $0.997\pm 0.002$ & $0.876\pm 0.068$ & $0.999\pm 5\mathrm{e}{-4}$ \\
\ExoMinerplusplusFFIVetter & $0.876\pm0.046$ & $0.852\pm0.030$ & $0.876\pm0.100$ & $0.996\pm0.004$ & $0.903\pm0.040$ & $0.999\pm1\mathrm{e}{-4}$ \\
\bottomrule
\end{tabular*}
\tablenotetext{a}{BD subclass has only 98 examples; therefore, recall values may not be statistically significant beyond two decimal places.}
\end{threeparttable}
\label{table:cv_subclass_recall}
\end{table*}

\section{Experiments}~\label{sec:experimental_setup}

In this section, we start by describing the setup used for running the experiments in this work. Specifically, we outline the data splits used for training and testing, the model training process, and the set of evaluation metrics used. We then detail the primary experiments conducted in this work, which include using both 2-minute and FFI data to train the models. The \ExoMiner\ code is publicly available on GitHub\footnote{\url{https://github.com/nasa/ExoMiner}}, and the preprocessed datasets of TESS SPOC 2-minute and FFI TCEs used in this work are available on Zenodo~\citep{martinho_2025_17707413}.

\subsection{Experimental Setup}

To evaluate the model under different data splits, we conducted 5-fold cross validation (CV) experiments, providing a more reliable estimate of performance across multiple splits. This approach ensures that all available data are used for both training and evaluation and helps assess generalization more robustly. The data are split into five folds such that the TCEs associated with the same target are included in the same fold. Given that all TCEs generated from the same target light curve and pixel data are either part of the training or test sets in any given CV iteration, we minimize data leakage. Furthermore, to ensure that the folds have approximately the same distribution of planets and non-planets as the overall dataset, we use a greedy strategy to assign TCEs to cross validation folds. For each target, all its TCEs are grouped together and assigned to the fold that holds the fewest planet TCEs. This iterative process ensures that the distribution of planet across folds remains as balanced as possible. 

Model performance is evaluated by using the following metrics: area under the precision-recall (PR AUC) and the receiver operating characteristic curves (ROC AUC), as well as precision, recall and accuracy at a classification threshold of $0.5$. Given that this work emphasizes the vetting of TCEs (in contrast to planet validation) and we aim to assess performance under stringent decision thresholds, we additionally report recall at the operating point where precision is 95\%, and precision at the operating point where recall is 95\%. The first metric indicates how many true planet TCEs are correctly identified when we require very few false positives (high precision), while the second shows how many predictions are correct when we aim to capture nearly all true planets (high recall). Together, these metrics illustrate the trade-off between completeness and reliability under strict classification criteria. 

 To evaluate and compare the two models, we report for each model the mean and standard deviation (SD) of the metrics across the five held-out test folds from cross validation. This provides a sense of the model’s stability across different partitions of the data. We compare the models based on mean$\pm$SD metrics across folds solely as a descriptive summary of their behavior, recognizing that these correlated estimates do not imply statistical significance, and should therefore be interpreted with caution.

For each cross validation iteration, we train a set of 10 models and construct an ensemble by averaging their prediction scores using the mean. This approach helps reduce variability caused by random weight initialization and the inherent stochasticity of mini-batch gradient descent during optimization. By training multiple models with different weight initializations, the ensemble captures variations in local minima encountered during the optimization of the loss function, resulting in a more accurate estimate of the model’s performance. 

All models are trained for 300 epochs, optimized using binary cross-entropy as the loss function and the Adam optimizer~\citep{adam2014method} ($\beta_1=0.900$, $\beta_2=0.999$, $\epsilon=1\mathrm{e}{-8}$). Model selection is performed using early stopping with a patience of 20 epochs to prevent overfitting; i.e.,\ if no improvement in the validation PR AUC is observed after 20 epochs, the training is stopped, and the model instance at that point is selected as the final model. The CV dataset can be found on Zenodo~\citep{martinho_2025_17707413}.

\subsection{Multi-source Learning}~\label{sec:multi-source-learning}

As the ongoing SPOC planet searches produce results for more sector runs, the number of FFI TCEs is expected to grow and surpass the 2-minute results. However, the current number of sector runs performed for the FFI data is significantly smaller than those for the 2-minute data, and only a single multisector run has been conducted as mentioned in Section~\ref{sec:sector_runs}. For this reason, using the 2-minute labeled dataset for training the model might result in improved performance on FFI data. Following the work in~\cite{Valizadegan_2025}, we conducted multi-source learning~\citep{zhuang2020comprehensive} using as sources the TESS SPOC 2-minute and FFI TCEs datasets for training. 

The FFI and 2-minute datasets differ in a few key important ways: 1) Cadence: the 2-minute data have a shorter cadence, whereas TESS full-frame images are sampled at either $30\ \text{min}$, $10\ \text{min}$, or $200\ \text{s}$ depending on the sector. The SPOC pipeline resamples the light curves to a common cadence of 10 minutes for the multisector runs since the S14--S55 search, partially reducing the difference; 2) Systematics correction: the 2-minute light curves benefit from more effective systematics correction, including less blending and crowding on average; 3) Vetting quality: the 2-minute data have higher-quality dispositions due to more extensive and focused vetting of transit signals. Although the 2-minute targets also appear in FFI data, the shorter and fewer sector runs for these data limit the detection of TCEs by the SPOC pipeline, particularly for longer-period TOIs. Additionally, targets observed exclusively in FFI data have received less scrutiny to date; and 4) Target population: FFI data cover a more diverse target population, including those targets already observed for 2-minute data as well as many more fainter targets.

Multi-source learning allows the model to simultaneously learn shared representations in a larger combined dataset that generalizes across both domains (i.e.,\ the two collection modes), which can be more effective than fine-tuning (e.g.,\ pre-training on the 2-minute dataset and then continuing training on the FFI dataset) since that may overwrite useful features learned from the 2-minute data. In this setting, the model is exposed during training to different noise levels and target populations which provides better generalization to new, unseen sectors.

\needspace{3\baselineskip} 
\subsection{Results and Discussion}

In this section, we start by presenting the results of the 5-fold CV experiments using \ExoMinerplusplus\ as the baseline study. Both \ExoMinerplusplus\ and \ExoMinerplusplusFFIVetter\ are evaluated on the same multi-source 5-fold CV dataset. We also investigate the change in performance between the two data sources.

\subsubsection{Cross Validation Results}


Table~\ref{table:cv_binary_results} presents the overall performance of the \ExoMinerplusplus\ and \ExoMinerplusplusFFIVetter\ models across a multi-source 5-fold cross validation procedure, evaluated using standard binary classification metrics. The first thing to notice is that the results for \ExoMinerplusplus\ are in line with those reported in~\cite{Valizadegan_2025} when the model is trained on TESS 2-minute data. Even though the dataset used in the current work has a different data distribution and includes more sector runs and FFI TCEs, the overall performance metrics remain consistent. 

\ExoMinerplusplus\ and \ExoMinerplusplusFFIVetter\ show small standard deviations, suggesting stable performance across folds. \ExoMinerplusplus\ shows higher standard deviation for recall at 95\% precision, indicating some operating-point sensitivity under fold changes. At this operating point, \ExoMinerplusplus\ achieves a recall of approximately 80\%, meaning that 95 out of 100 predicted planets are true exoplanets, and 80 out of 100 exoplanets are retrieved. As an example, if we assume a catalog of 1,000 objects following the same ratio of planet TCEs to non-planet TCEs as our combined 2-min/FFI dataset (approximately 5.6\% planet TCEs vs. 94.4\% non-planet TCEs), then at an operating point of 95\% precision and 80\% recall, the model would predict 47 TCEs as planets, of which 2 are non-planets. Furthermore, the model would miss 11 planet TCEs out of the 56 present in the catalog.

Table~\ref{table:cv_subclass_recall} details the recall values for each subclass at a fixed classification threshold of 0.5, highlighting the models’ ability to correctly identify positive instances within each specific subclass. Across the five folds, these results combined with the ones shown in Table~\ref{table:cv_binary_results} suggest that \ExoMinerplusplus\ has higher recall, accuracy, PR AUC, and ROC AUC, whereas \ExoMinerplusplusFFIVetter\ shows stronger performance on specific negative subclasses (FP, BD). Overall, \ExoMinerplusplus\ offers stronger global performance and recall, whereas \ExoMinerplusplusFFIVetter\ provides targeted improvements in specific negative subclasses.

\begin{figure}[htb!]
\vskip 0.1in
\begin{center}
\centerline{\includegraphics[width=\columnwidth]{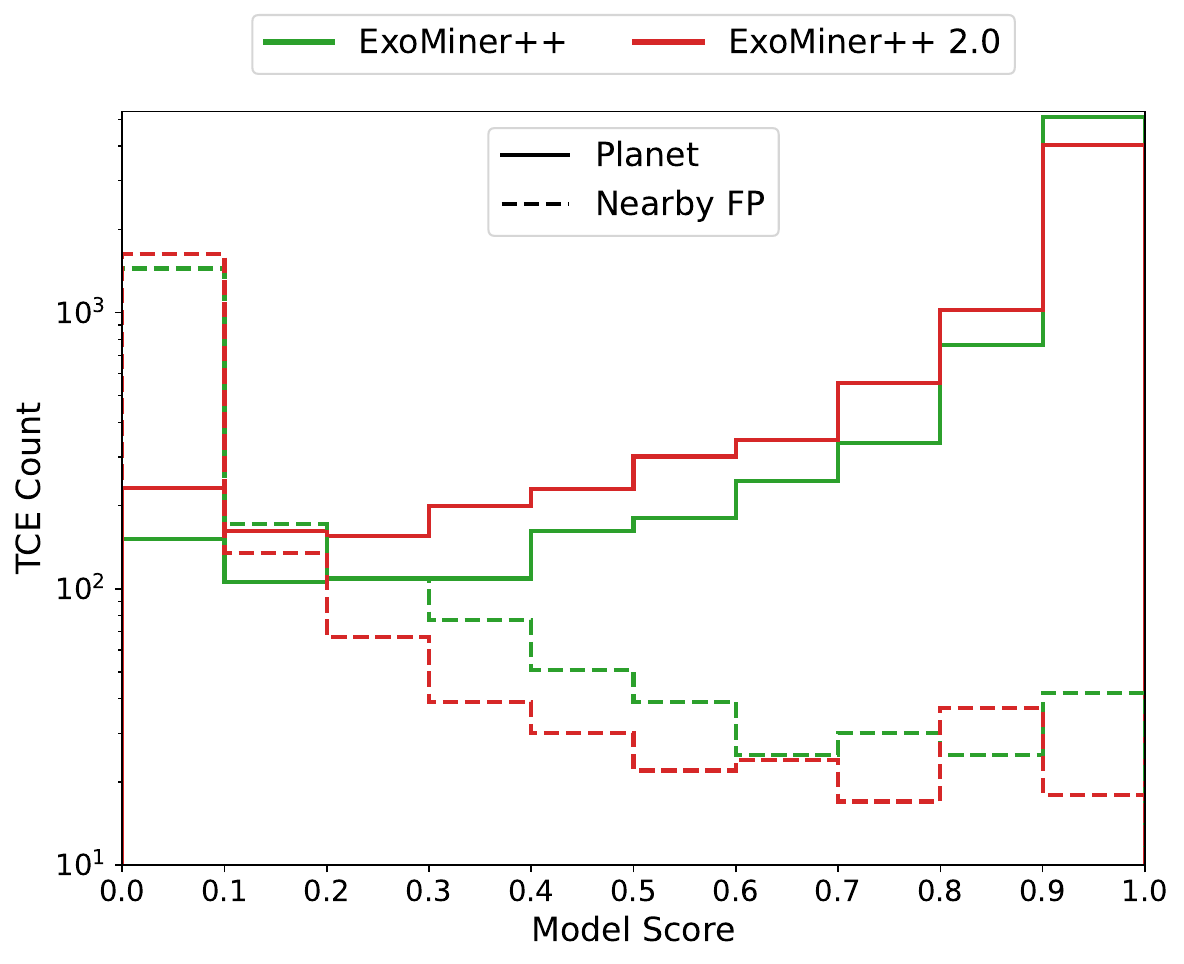}}
\caption{Distribution of \ExoMinerplusplus\ and \ExoMinerplusplusFFIVetter\ scores for planet and nearby false positive TCEs.}
\label{fig:hist_model_scores_nearby-fp-vs-planets_compare-experiments}
\end{center}
\vskip -0.3in
\end{figure}

Since we aim to assess whether incorporating neighbors' data improves the model's ability to reduce confidence for TCEs with observable transit-source offsets near bright stars, we also examined the recall of both models for nearby false positives, specifically TOIs classified as nearby planet candidates and eclipsing binaries in SG1. The recall for these nearby FPs is 0.92 and 0.94 for \ExoMinerplusplus\ and \ExoMinerplusplusFFIVetter, respectively. Figure~\ref{fig:hist_model_scores_nearby-fp-vs-planets_compare-experiments} shows the distribution of scores for the two models for both planet and nearby FP TCEs. The shift in both distributions indicates that \ExoMinerplusplusFFIVetter\ lowered the scores of all objects, true planets or background transits, thus confirming that this model is more conservative than \ExoMinerplusplus.

\begin{figure}[htb!]
\vskip 0.1in
\begin{center}
\centerline{\includegraphics[width=\columnwidth]{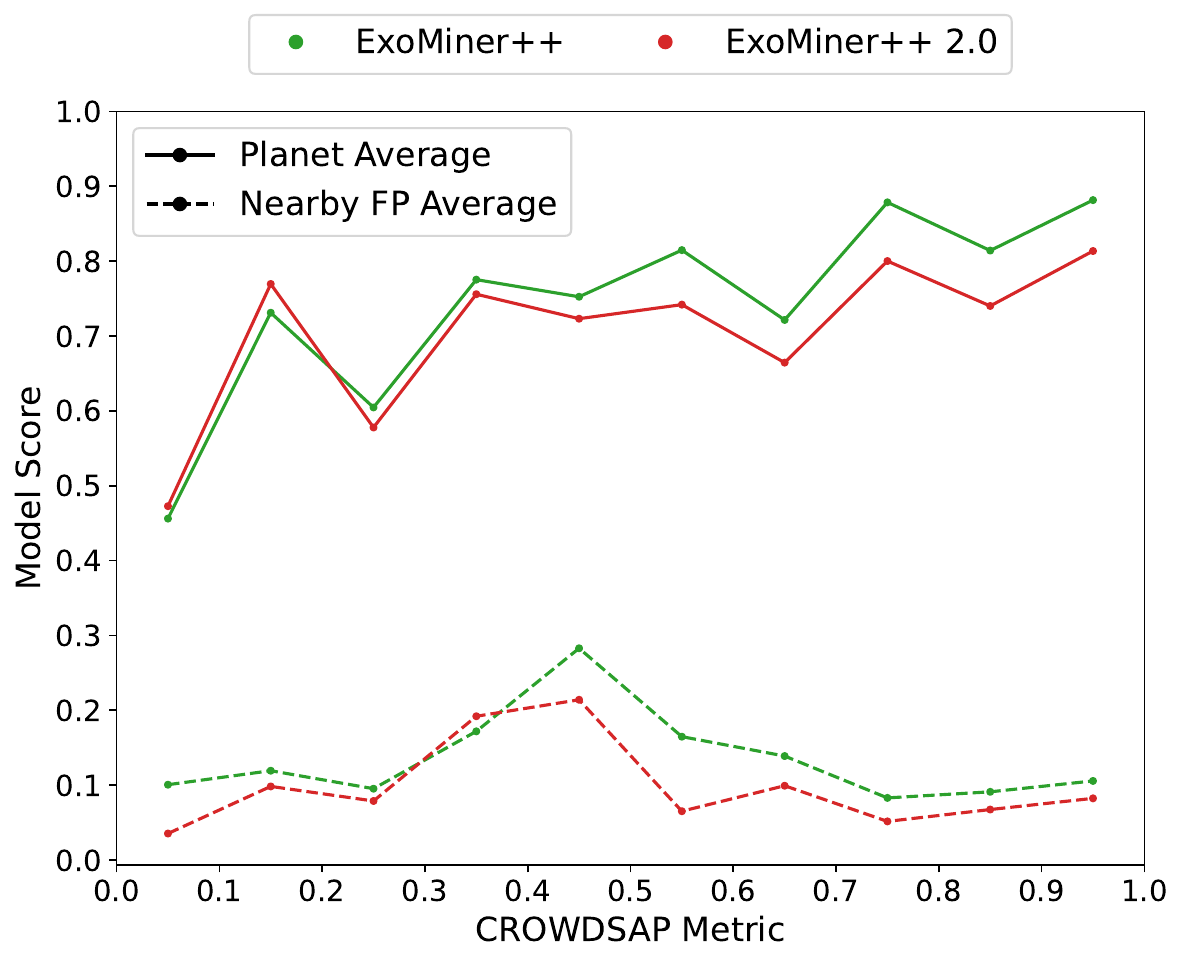}}
\caption{Average \ExoMinerplusplus\ and \ExoMinerplusplusFFIVetter\ score for planet and nearby false positive TCEs as a function of CROWDSAP metric.}
\label{fig:scatter-plot_crowdsap_vs_mean-score_planets_vs_nearby-fps}
\end{center}
\vskip -0.3in
\end{figure}

A similar behavior is also observed when we plot the mean average score for planets and nearby FPs as a function of the CROWDSAP metric (Figure~\ref{fig:scatter-plot_crowdsap_vs_mean-score_planets_vs_nearby-fps}). TCEs observed in multiple sectors were excluded from this analysis because, in those cases, there are multiple CROWDSAP metrics available, one for each sector where the target was observed, and there is no straightforward method to aggregate these metric values into a single statistic. The CROWDSAP metric in TESS data quantifies the fraction of flux in the photometric aperture that originates from the target star; values near unity indicate minimal contamination, whereas lower values indicate significant flux contribution from nearby stars or background sources. As the contamination increases (i.e.,\ lower CROWDSAP values), both models exhibit a decrease in the planet average score, while the average score for nearby FPs remains consistent (results below CROWDSAP $\approx0.5$ are less reliable given that those bins contain fewer than 100 TCEs). However, even in cases where nearly all of the flux in the aperture comes from the target star (i.e.,\ CROWDSAP close to 1), \ExoMinerplusplusFFIVetter\ still shows an average lower score for both planet and nearby FP TCEs. This conservativeness translates into a tradeoff, as the model degrades its performance on planet TCEs to correctly classify nearby false positives. Three hypotheses can explain this: 1) the large pixel scale ($21\arcsec$) means that even in moderately crowded fields, the light from the target star is often mixed with nearby stars that fall into the same photometric aperture or even the same pixel and can still lead to errors in the transit source location estimation even when conducting difference image analysis. This limitation constitutes a lower bound that ultimately limits the capability of the model to separate on-target transits from background sources; 2) the small number of nearby FPs relative to the size of the dataset, combined with the joint optimization of the model means that it is challenging for the model to optimize its `Difference Image' branch to learn features useful for detecting real transit source offsets. For example, pretraining the `Difference Image' branch on a dedicated task that distinguishes on-target transit signals from those originating in nearby stars could enable the branch to learn discriminative representations for this scenario; 3) no dedicated hyperparameter optimization (HPO) was performed for this architecture, particularly the `Difference Image' branch. We expect that by conducting a new HPO run we can find a more optimal architecture that can use the difference image data more effectively, including the neighbors information. Other potential future improvements to enhance the performance of the `Difference Image' branch in cases where the transit source may be a neighboring star include incorporating a lightweight spatial attention mechanism inspired by~\citep{woo2018cbam} to guide the model toward salient regions of the image such as the transit signal location, while suppressing irrelevant areas like pixels without known stars. Another promising direction is to select difference images from sectors with higher quality metrics rather than relying on random sampling with replacement, ensuring the model is exposed to a set of higher-quality inputs.

\subsubsection{Comparison Between Data Sources}

As outlined in Section~\ref{sec:multi-source-learning}, the 2-minute and FFI datasets exhibit distinct observational and noise characteristics, which can influence model behavior. In this section, we start by comparing the overall performance on these two datasets. To remove confounding effects from this analysis, we exclude NTP TCEs since this subset is dependent on the availability of TEC's flux triage results. As mentioned in Section~\ref{sec:label-assignment}, we had access to an inconsistent set of these results: for 2-minute data we used the flux triage results for single-sector runs S1 through S41, while for FFI we had results for single-sector runs S40 through S73. Besides not having the results for the same sector runs, the numbers of NTP TCEs generated by these two results varies significantly as shown in Tables~\ref{table:2min_counts_dataset} and~\ref{table:ffi_counts_dataset}. Furthermore, the SPOC planet searches for FFI data start only with Sector 36, and so far only one multisector run was conducted. Because of this, the number of multisector runs from 2-minute data is significantly larger, and on average longer than the single multisector run conducted for FFI data (S56--S69). With this bias in mind, for this analysis we considered only TCEs from the planet searches that use the same sectors, which in this case are all single-sector runs from Sector 36 through Sector 72, totaling 37 runs. Given that the ephemeris matching was conducted for the same sector runs and using the same label sources, the only differences in these two sets come from the target selection criteria for FFI and 2-minute data, and the operation of the SPOC pipeline when performing planet searches in these two data collection modes. Table~\ref{table:dataset_results_2min_vs_ffi} shows the number of TCEs, targets, and events (i.e.,\ TOIs and Prsa's EBs) for 2-minute and FFI data after applying these filters. 

\begin{table}[t]
\centering
\caption{Comparison between 2-minute and FFI datasets. NTP TCEs were excluded and only TCEs from the same sector runs were considered from both datasets.}
\small
\begin{tabularx}{\columnwidth}{@{\extracolsep{\fill}} lcc}
\toprule
Category & 2-minute & FFI \\
\midrule
TCEs & 6,020 & 6,419 \\
Targets & 2,060 & 2,226 \\
Events & 2,096 & 2,241 \\
KP TCEs (\%) & 801 (13.31\%) & 604 (9.41\%) \\
CP TCEs (\%) & 669 (11.11\%) & 776 (12.09\%) \\
BD TCEs (\%) & 13 (0.22\%) & 16 (0.25\%) \\
EB TCEs (\%) & 3,747 (62.24\%) & 4,051 (63.11\%) \\
FP TCEs (\%) & 790 (13.12\%) & 972 (15.14\%) \\
\bottomrule
\end{tabularx}
\label{table:dataset_results_2min_vs_ffi}
\end{table}

As expected, the FFI data contain more TCEs as the number of targets selected for FFI planet searches was larger than for 2-min. Following this trend, the number of events matched to TCEs was also larger. However, both datasets have targets and events that are exclusive to each other. In the case of events, the number of these objects that are shared between 2-minute and FFI datasets is 1,708, while 533 and 388 are found exclusively in FFI and 2-minute data, respectively. Interestingly, the number of detected KP TCEs in 2-minute data is significantly larger than for FFI data, while a similar trend is shown for CP TCEs but favoring FFI data. Besides those differences, the FFI dataset also shows significantly more FP TCEs detected than the 2-minute counterpart.

\begin{table*}[!t]
\footnotesize
\centering
\caption{Binary classification performance for \ExoMinerplusplusFFIVetter\ from CV experiment computed separately for the 2-minute and FFI datasets. Precision, recall, and accuracy are computed at a classification threshold of 0.5. The best performer is highlighted in bold. Point estimates are computed across the CV iterations by aggregating all test sets.}
\label{table:cv_binary_results_2min_vs_ffi}

\setlength{\tabcolsep}{4pt} 
\renewcommand{\arraystretch}{1.2} 
\begin{tabular*}{\textwidth}{@{\extracolsep{\fill}} l | >{\centering\arraybackslash}p{2.5cm} | >{\centering\arraybackslash}p{2.0cm} | >{\centering\arraybackslash}p{2.5cm} | >{\centering\arraybackslash}p{2.0cm} | >{\centering\arraybackslash}p{1.5cm} | >{\centering\arraybackslash}p{2.5cm} | >{\centering\arraybackslash}p{2.0cm} @{}}
\hline
Source & Precision \& Recall\tablenotemark{a} & PR AUC\tablenotemark{a} & Normalized PR AUC & ROC AUC & Accuracy\tablenotemark{a} & Precision @ 95\% Recall\tablenotemark{a} & Recall @ 95\% Precision\tablenotemark{a} \\
\hline
2-minute & \textbf{0.937} \& 0.845 & \textbf{0.967} & \textbf{0.957} & \textbf{0.989} & 0.948 & \textbf{0.876} & \textbf{0.767} \\
FFI & 0.917 \& \textbf{0.854} & 0.952 & 0.938 & 0.987 & \textbf{0.952} & 0.854 & 0.651 \\
\hline
\end{tabular*}
\tablenotetext{a}{These metrics should be compared with caution because they are sensitive to class distribution differences between datasets.}
\end{table*}

Table~\ref{table:cv_binary_results_2min_vs_ffi} shows the performance metrics for \ExoMinerplusplusFFIVetter\ on these two datasets. The metrics were computed by aggregating the test set predictions across all CV iterations. Overall, the 2-minute dataset shows higher performance, including on normalized PR AUC and ROC AUC, both of which are considered less sensitive to variations in class distribution across the two datasets. The normalized PR AUC is computed as:
\begin{align}
\text{Normalized PR AUC} &= \frac{\text{PR AUC} - \gamma}{1 - \gamma}, \\
\gamma &= \frac{N_{\text{pos}}}{N_{\text{total}}},
\end{align}
where $\gamma$ denotes the prevalence of positive examples in the dataset. This normalization maps the random classifier baseline to zero and a perfect classifier to 1.

\begin{figure}{}
  \centering
    \subfigure[Target Count]{%
        \label{fig:hist_targets_datasets_ffi_vs_2min_crowdsap-counts}%
        \includegraphics[width=\linewidth]{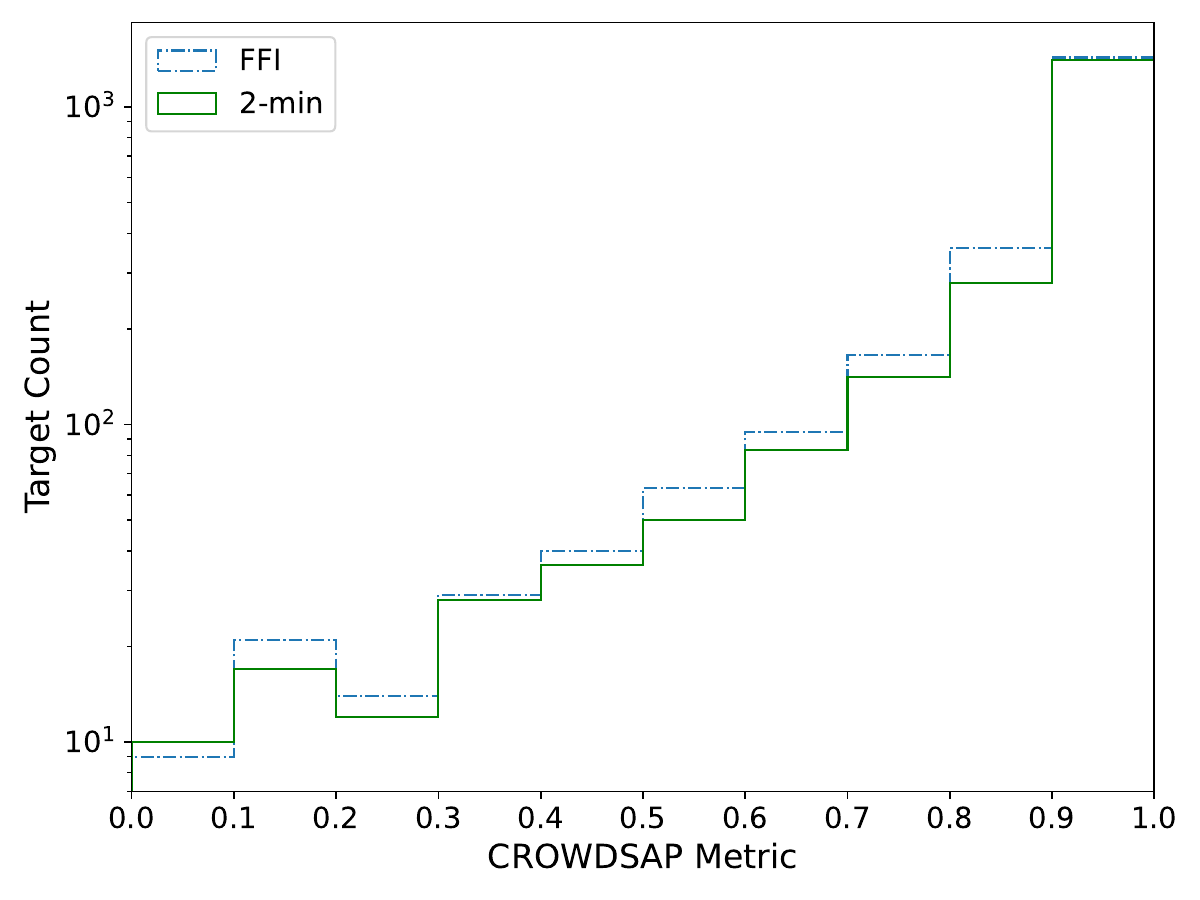}}\\[1ex]
    \subfigure[Target Density]{%
        \label{fig:hist_targets_datasets_ffi_vs_2min_crowdsap-density}%
        \includegraphics[width=\linewidth]{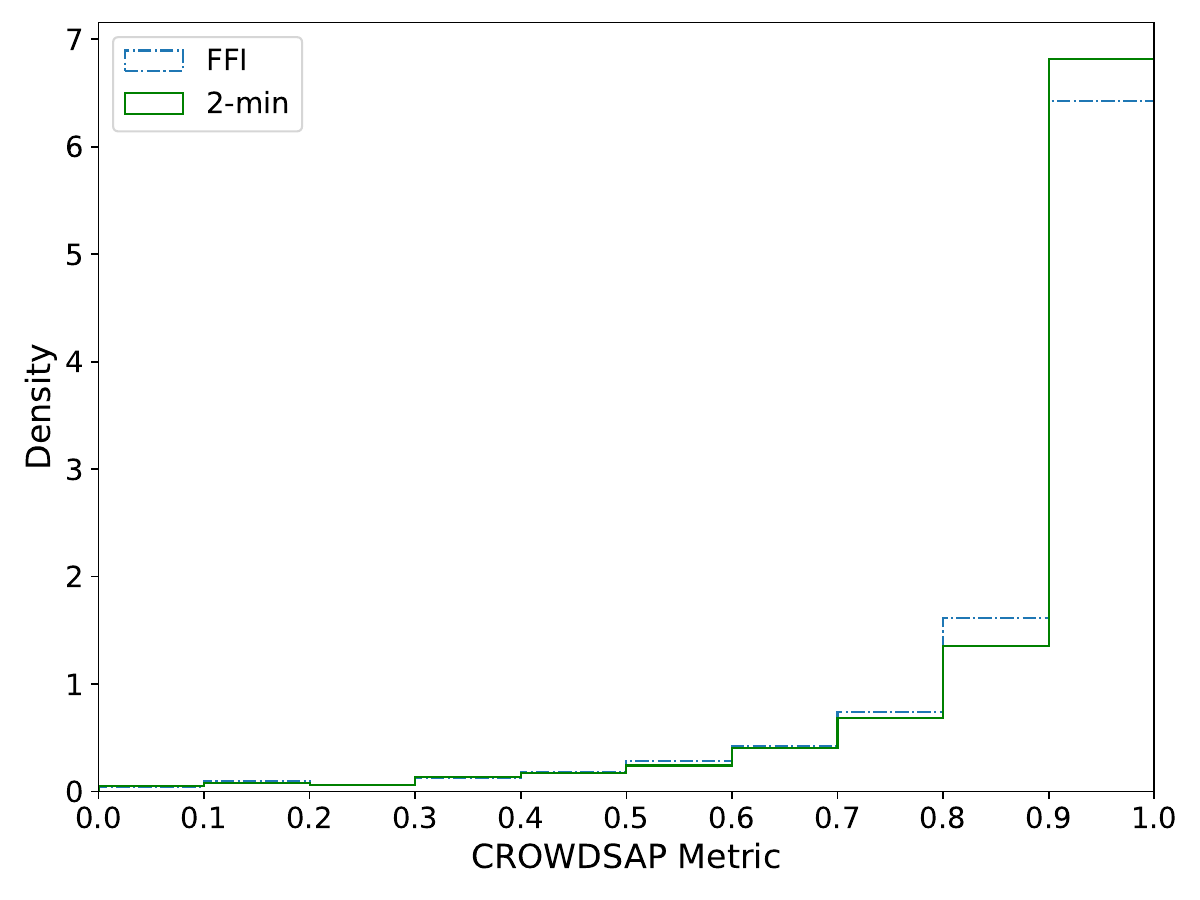}}
\caption{Target population for the 2-minute and FFI TCE datasets as a function of CROWDSAP Metric.}
\label{fig:hist_targets_2min_vs_ffi_crowdsap}
\end{figure}

To investigate the causes behind the lower performance in the FFI dataset, we examined parameters that could explain these differences. More concretely, we analyzed the distribution of targets as function of $T_{mag}$ and Gaia RUWE, as well as the distribution of TCEs as function of their multiple event statistic (MES) and TCE model SNR. None of these parameters showed a clear shift in distribution between the two datasets. However, the distribution of the CROWDSAP metric revealed a noticeable shift for FFI targets toward lower values (Figure~\ref{fig:hist_targets_2min_vs_ffi_crowdsap}). A non-negligible fraction of FFI targets exhibit CROWDSAP values below 0.9, meaning that at least 10\% of the flux in the aperture originates from other sources. Such contamination can dilute transit depths, bias planet radius estimates, reduce MES and SNR, and introduce systematics that are more difficult to correct by algorithms such as the PDC module in the SPOC pipeline. Furthermore, blended eclipsing binaries and variable stars within the aperture can mimic transit-like signals, increasing the false positive rate. Although MES and SNR distributions do not appear strongly affected, the deterioration in \ExoMinerplusplusFFIVetter\ performance suggests that contamination impacts other classification-relevant features, such as detrending quality.

\begin{table}[t]
\footnotesize
\caption{Scores and dispositions of \ExoMinerplusplusFFIVetter\ model for unlabeled FFI SPOC TCEs. This table describes the available columns. The full table is available online.}
\label{table:vetting-catalog}
\begin{tabularx}{\columnwidth}{@{}p{0.2\columnwidth}>{\raggedright\arraybackslash}X@{}}
\hline
\textbf{Column} & \textbf{Description} \\
\hline
uid & Unique TCE ID that includes TIC ID, planet number, and sector run in the format tic\_id-spoc\_tce\_planet\_number-Ssector\_run \\
target\_id & TIC ID \\
tce\_plnt\_num & SPOC TCE planet number \\
sector\_run & SPOC sector run \\
tce\_period & TCE period (days) \\
tce\_duration & TCE duration (hours) \\
tce\_time0bt & TCE epoch (BTJD) \\
tce\_depth & TCE depth (ppm) \\
tce\_prad & TCE planet radius (Earth radii) \\
mes & TCE MES \\
tce\_model\_snr & TCE model SNR \\
tmag & TESS magnitude \\
ruwe & Gaia DR2 RUWE \\
n\_tois\_in\_tic & Number of known TOIs in the TIC\tablenotemark{a} \\
tois\_in\_tic & List of known TOIs in the TIC\tablenotemark{a} \\
DV full report & URL to the DV full report in MAST \\
DV summary report & URL to the DV summary report in MAST \\
DV mini report & URL to the DV mini report in MAST \\
score\_cv\_iter\_i & For $i \in [0,5]$, score of the \ExoMinerplusplusFFIVetter\ model trained on CV fold $i$ \\
mean\_score & Average score of 5 \ExoMinerplusplusFFIVetter\ models \\
std\_score & Standard deviation of the scores of 5 \ExoMinerplusplusFFIVetter\ models \\
model\_label & `PC' if \ExoMinerplusplusFFIVetter\ score $> 0.5$, `FP' otherwise \\
\hline
\end{tabularx}
\tablenotetext{a}{As of September 22, 2025.}
\end{table}

\section{Vetting FFI TCEs}~\label{sec:vetting_catalog}


To vet unlabeled FFI TCEs, we employed the ensemble of trained models from the \ExoMinerplusplusFFIVetter\ CV experiment. Although training a single final model using the same setup would simplify deployment, it would likely be less robust than an ensemble trained on different subsets of the data. Leveraging the CV ensemble improves robustness and reduces variance in predictions for the unlabeled dataset. For each TCE, we aggregate predictions from the five CV folds to compute the mean and standard deviation, providing both a ranking metric and an uncertainty estimate. The mean prediction can be used to prioritize candidates, while the standard deviation reflects model disagreement. High uncertainty indicates ambiguous signals, which may correspond to borderline cases and unusual morphologies, and thus making them scientifically interesting.


The complete table of predictions for UNK TCEs is available on Zenodo~\citep{martinho_2025_17707413}. Table~\ref{table:vetting-catalog} describes the contents of this table, which lists unlabeled TCEs ranked by decreasing prediction score. In addition to TCEs corresponding to TOIs with labels we treat as uncertain (e.g., PC), the table also contains some TCEs associated with TOIs that we treat as certain (e.g., KP). These cases arise for several reasons: (1) failure to match a TCE to its corresponding TOI due to uncertainties in the estimated ephemerides and the propagation of errors. The matching procedure relies on period and epoch estimates, and significant uncertainties, particularly when epochs are widely separated, can prevent proper alignment of transit events. This issue is more common for single-sector and longer-period TCEs, where fewer observed transits lead to poorer ephemeris constraints; (2) failure to match a TCE to a TOI because the period detected by the SPOC pipeline is a multiple or fraction of the object’s true period; and (3) promotion of objects to TOIs after the ExoFOP TOI catalog used in this work was downloaded.

\begin{figure}[htb!]
\vskip 0.1in
\begin{center}
\centerline{\includegraphics[width=\columnwidth]{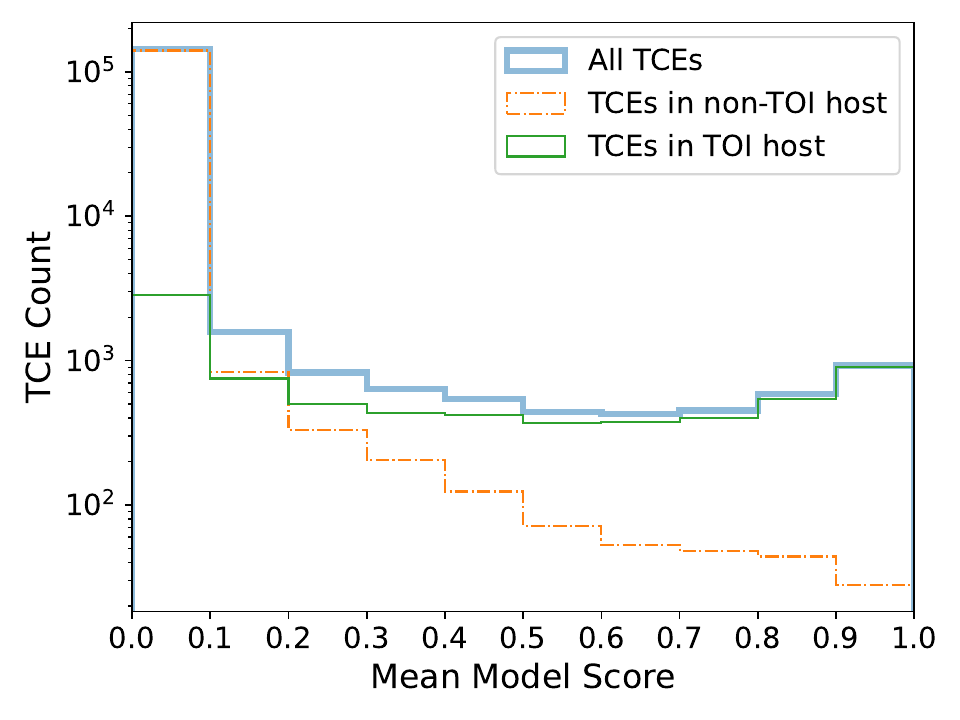}}
\caption{Distribution of \ExoMinerplusplusFFIVetter\ mean scores for unlabeled FFI SPOC TCEs.}
\label{fig:hist_scores_unk-ffi-tces}
\end{center}
\vskip -0.3in
\end{figure}

The predicted mean score distribution for the unlabeled FFI TCEs is presented in Figure~\ref{fig:hist_scores_unk-ffi-tces}, which also distinguishes between candidates orbiting targets with known TOIs and those orbiting stars with no known TOIs. Because they are less likely to be previously cataloged, the set of TCEs associated with targets without known TOIs is considered a higher priority for further vetting. To quantify the TCEs in the full set of unlabeled examples, we find that 2,831 TCEs score above 0.5, 926 above 0.9, and 447 above 0.95, while none scored above 0.99. 

\begin{figure}[htb!]
\vskip 0.1in
\begin{center}
\centerline{\includegraphics[width=\columnwidth]{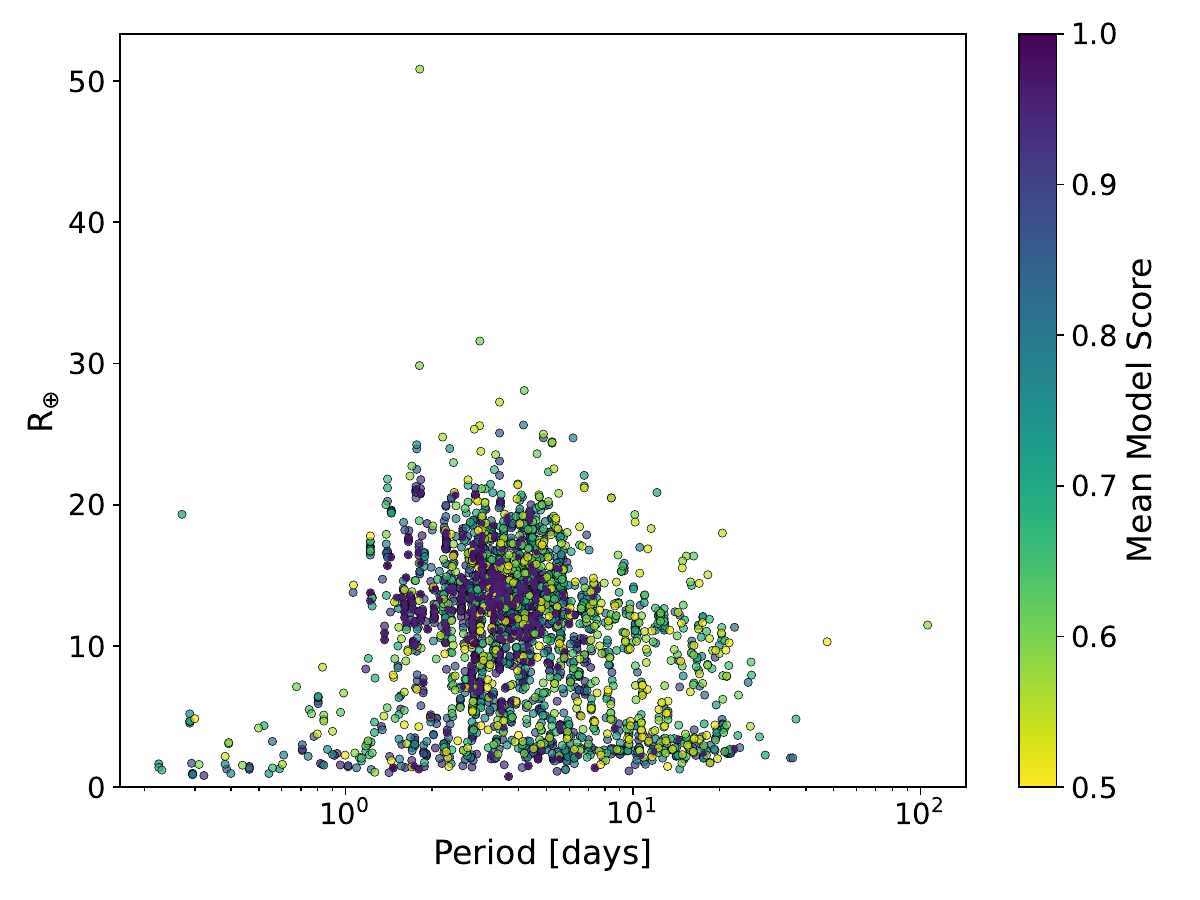}}
\caption{Distribution of SPOC DV estimated orbital period versus planet radius for UNK FFI TCEs with \ExoMinerplusplusFFIVetter\ mean scores greater than 0.5. Each data point represents a TCE and is colored based on the assigned prediction score.}
\label{fig:scatter_period_radius_scores_unk-ffi-tces_vet-thr-0.5}
\end{center}
\vskip -0.3in
\end{figure}

Figure~\ref{fig:scatter_period_radius_scores_unk-ffi-tces_vet-thr-0.5} shows the distribution of the orbital period against planet radius for TCEs whose scores were above 0.5. As expected, most TCEs have an orbital period shorter than the duration of a single sector. This occurs because the FFI dataset currently lacks multisector runs and yields on-average lower SNR for longer-period TCEs (i.e., period $\gtrsim27$ days) due to the fewer observations available which in turn lowers the model's predicted scores for those examples. The longest-period TCE ($\approx 106$ days) has a mean score of about 0.59 and an estimated planet radius of approximately $11.5\ R_{\oplus}$. A cluster of TCEs with scores greater than 0.9 is visible for orbital periods in the range of \text{$[2,5]$} days and planet radius in the range of \text{$[10,20]\ R_{\oplus}$}, consistent with the expected parameters for hot Jupiters. For the 58 TCEs in the range of potential terrestrial planets ($R_{\oplus}<1.7$), there is no clear pattern in their distribution of period and planet radius.



\section{Limitations}

\ExoMinerplusplusFFIVetter\ demonstrates competent vetting capability on FFI data, but it remains subject to several practical limitations. First, the 21\arcsec\ TESS pixel scale limits the ability to distinguish on‑target from nearby contaminating sources, even with increased image resolution for the encoding of neighboring stars. Second, FFI light curves exhibit greater variability and fewer multisector runs than 2‑minute data, reducing diagnostic richness and contributing to lower overall performance on FFI TCEs. Third, the labeled FFI set remains relatively small, particularly for rare subclasses, limiting the robustness of subclass‑specific recall estimates. Finally, although the architecture introduces multiple enhancements, it has not undergone a dedicated hyperparameter optimization, and the `Difference Image' branch likely has room for further refinement.

\section{Conclusions}

In this work, we extended the \ExoMinerplusplus\ framework to vet transit signals detected in TESS FFI data, introducing architectural, preprocessing, and data‑integration improvements tailored to the unique characteristics of more contamination‑prone FFI observations. By leveraging multi‑source learning, incorporating neighboring star information, and redesigning model components, \ExoMinerplusplusFFIVetter\ achieves effective planet versus non‑planet discrimination across both FFI and 2‑minute datasets. The resulting vetted catalog of unlabeled FFI TCEs expands the accessible parameter space for TESS transit searches and provides a valuable resource for prioritizing follow‑up observations. Key contributions of this work include:

\begin{itemize}

\item First incorporation of neighboring‑star information into a TESS machine‑learning vetter. We introduce an image‑based representation encoding the relative brightness and location of known TIC neighbors, enabling the model to reason about off‑target transit sources.

\item Architectural improvements to \ExoMinerplusplus.
We merge transit-view branches, remove `Momentum Dump' branch, add global average pooling for parameter efficiency, introduce learned downsampling, incorporate batch and layer normalization, and strengthen residual connections for more stable training.

\item Multi‑source learning across 2‑minute and FFI datasets. The model jointly learns from both domains, improving FFI performance by leveraging the richer labels and cleaner systematics of the 2‑minute dataset.

\item A vetted catalog of 150k previously unlabeled FFI TCEs. We generate mean predictions and uncertainty estimates from an ensemble of CV models, offering a ranked list of candidates for community follow‑up, population studies, and TOI augmentation.

\item Improved sensitivity to off‑target and contamination‑driven false positives. \ExoMinerplusplusFFIVetter\ systematically reduces the scores of transit‑like signals originating near bright neighbors, making it more conservative in crowded fields.

\end{itemize}

Together, these advancements establish \ExoMinerplusplusFFIVetter\ as a robust, scalable, and contamination‑aware vetting system for TESS FFI transit searches. As additional FFI sectors become available, we expect further improvements through expanded multi‑source training, dedicated hyperparameter optimization of the `Difference Image' branch, and the incorporation of attention mechanisms.

\section*{Acknowledgments}

MM and HV are supported through TESS XRP 2022 contract 22-XRP22\_2-0173, NASA Academic Mission Services (NAMS) contract number NNA16BD14C as well as the Intelligent Systems Research and Development Support-3 (ISRDS-3) Contract 80ARC020D00100. DC, JT, MJ, and BT are supported through NASA Cooperative Agreement 80NSSC21M0079.

Resources supporting this work were provided by the NASA High-End Computing (HEC) Program through the NASA Advanced Supercomputing (NAS) Division at Ames Research Center for the production of the Kepler SOC and the TESS SPOC data products and for training our deep learning model and preprocessing the data to be ingested by the model. 

Funding for the TESS mission is provided by NASA's Science Mission Directorate. This paper made use of data collected by the TESS mission and are publicly available from the Mikulski Archive for Space Telescopes (MAST) operated by the Space Telescope Science Institute (STScI). We acknowledge the use of public TESS data from pipelines at the TESS Science Office and at the TESS Science Processing Operations Center. This research has made use of the Exoplanet Follow-up Observation Program (ExoFOP; DOI: 10.26134/ExoFOP5) website, which is operated by the California Institute of Technology, under contract with the National Aeronautics and Space Administration under the Exoplanet Exploration Program. We would like to thank ExoFOP-TESS for hosting and sharing vetted TOI information and follow-up results, and the TESS Science Office and the TFOP SG1 Working Group, led by Dr. Karen A. Collins (CfA/SAO), for their vetting and ground-based follow-up of TESS Objects of Interest.

This material is based upon work supported by the NASA under Agreement No.\ 80NSSC21K0593 for the program ``Alien Earths''. The results reported herein benefited from collaborations and/or information exchange within NASA’s Nexus for Exoplanet System Science (NExSS) research coordination network sponsored by NASA’s Science Mission Directorate.

We would also like to acknowledge the use of Microsoft Enterprise Copilot~\citep{microsoft2025copilot}, an AI-powered writing assistant, in editing portions of this manuscript to improve clarity and writing quality.

\bibliography{AI_bib,ExoPlanet}{}




\appendix

\section{Details of Architecture for \ExoMinerplusplusFFIVetter}
\label{sec:optimized_exominer}


We provide in detail the architecture and optimization parameters of \ExoMinerplusplusFFIVetter:
\begin{itemize}

    \item All convolutional layers use a stride of 1, and their weights are initialized randomly following He initialization~\citep{he2015delving}.
    \item All convolutional layers use `same' padding (i.e.,\ the feature map size is preserved), except for the first layer in each convolutional block, whose padding was set to `valid' (thus downsampling the feature map similar to pooling). The only exception is the `Difference Image' branch, whose convolutional layers all apply `same' padding.
    \item All convolutional and FC layers are followed by a parametric rectified linear unit activation~\citep[pReLU]{he2015delving}.
    \item All skip connections apply a convolutional layer with the same number of kernels, type of padding, and kernel size and stride as the first convolutional layer in each convolutional block so the feature map dimensions match. The kernel is initialized so the convolutional layer works initially as an identity layer.
    \item Layer normalization layers were applied before the classification head and after each FC layer in the classification head. All layer normalization layers use $\epsilon=1\mathrm{e}{-3}$, with center and scale parameters initialized to `zeros' and `ones', respectively.
    \item All batch normalization layers use $\mu=0.99$ and $\epsilon=1\mathrm{e}{-3}$, with center and scale parameters initialized to `zeros' and `ones', respectively.
    \item All FC layers in the classification head apply a dropout with rate of $2.15\mathrm{e}{-2}$, and their weights are initialized randomly following Glorot uniform initialization~\citep{glorot2010understanding}.
    \item Batch size was set to 256 to provide stable statistics for batch normalization.
    \item Learning rate was set to $4.18\mathrm{e}{-5}$ to compensate batch size increase.
\end{itemize}

\end{document}